# Near-Zero Crosstalk and Ultra-Low Loss Waveguide Crossings Enabled by three-dimensional $Ta_2O_5$-on-LNOI Integrated Photonic Platform

Boyang Nan[1,2,§], Yuan Ren[1,2,§], Rongbo Wu[2,*], Lvbin Song[1,2], Ruixue Liu[1,2], Yong Zheng[2], Min Wang[2,*] and Ya Cheng[1,2,3,4,5,6,7,*]

[1]State Key Laboratory of Precision Spectroscopy, East China Normal University, Shanghai 200062, China
[2]The Extreme Optoelectromechanics Laboratory (XXL), School of Physics and Electronic Science, East China Normal University, Shanghai 200241, China
[3]State Key Laboratory of High Field Laser Physics and CAS Center for Excellence in Ultra-intense Laser Science, Shanghai Institute of Optics and Fine Mechanics (SIOM), Chinese Academy of Sciences (CAS), Shanghai 201800, China
[4]Collaborative Innovation Center of Extreme Optics, Shanxi University, Taiyuan 030006, China
[5]Collaborative Innovation Center of Light Manipulations and Applications, Shandong Normal University, Jinan 250358, China
[6]Shanghai Research Center for Quantum Sciences, Shanghai 201315, China
[7]Hefei National Laboratory, Hefei 230088, China
§These authors contributed equally to this work and should be considered co-first-author
*Authors to whom any correspondence should be addressed

E-mail: rbwu@phy.ecnu.edu.cn, mwang@phy.ecnu.edu.cn and ycheng@phy.ecnu.edu.cn

## Abstract

Waveguide crossings represent one of the most critical components in very-large-scale photonic integration (VLSPI). Three-dimensional waveguide crossings, which distribute optical pathways across multiple planes, can achieve near-zero crosstalk and extremely low crossing-induced loss. However, they face an intrinsic trade-off between interlayer crossing performance and coupling efficiency. To address this challenge, we developed a low-cost fabrication method for 3D waveguide crossings by exploiting the edge rounding effect inherent to chemical mechanical polishing (CMP). Using this method, we demonstrate waveguide crossings with average loss below 0.002 dB and crosstalk below -62 dB on $Ta_2O_5$-on-LNOI integrated photonic platform. Our method maintains full compatibility with conventional semiconductor manufacturing technology and paves the way for realizing VLSPI on the thin-film lithium niobate platform.

**Keywords**：Waveguide crossing, Insertion loss, Crosstalk, Thin-film lithium niobate, Tantalum oxide

## Highlights

1. By integrating PLACE technology with dry etching processes, a novel high-performance three-dimensional waveguide cross-interconnect architecture is proposed.

2. By leveraging the inherent rounding effect of chemical mechanical polishing, efficient optical coupling of vertically stacked waveguides is achieved, enabling photons to propagate with minimal loss on the $Ta_2O_5$-on-LNOI integrated platform.
3. Achieve ultra-low insertion loss (< 0.002 dB) and near-zero crosstalk (< 62 dB), along with large-scale integration (300 cascades).

## 1. Introduction

Photonic integrated circuits (PICs), enabled by their inherent advantages, including large bandwidth[1], low latency[2], high parallelism[3], and energy efficiency[4], have become a promising complement to integrated electronics, significantly expanding the frontiers of communication and computing systems[5]. Waveguide crossings allow optical waveguides to cross over each other with minimal crosstalk and loss, facilitating flexible reconfiguration of waveguide array architectures[6-9]. This feature is becoming increasingly indispensable as the scale and complexity of PICs continue to grow. Multimode interference (MMI)-based designs have become the most widely adopted solution to realize in-plane waveguide crossings owing to their design simplicity and fabrication robustness[10,11]. While these waveguide crossings exhibit relatively low insertion loss (typically around 0.1 dB) and crosstalk (typically -50 dB) at the individual device level, these performance metrics become inadequate for large-scale PICs that require hundreds or even thousands of cascaded crossings, where crosstalk and losses accumulate[12,13]. The pursuit of minimized loss and crosstalk has led to the development of three-dimensional (3D) waveguide crossings[14-16], which utilize vertical interlayer couplers with low insertion loss to distribute light across multiple planes. This approach fundamentally prevents waveguide intersection in the same plane, and by virtue of the physical separation between light in different layers, achieves intrinsic near-zero crosstalk and ultra-low crossing-induced loss[17]. However, an inherent trade-off exists between interlayer crossing performance and coupling efficiency. While increased layer spacing can reduce crossing-induced losses, this comes at the expense of elevated transition losses during interlayer coupling[18]. Various coupler architectures have been investigated to overcome this trade-off, including grating couplers and triple-layer couplers. Grating couplers[19,20] offer compact footprints and enable optical coupling between two waveguide layers with a large spacing, but suffer from relatively high insertion loss and narrow operational bandwidth. In contrast, triple-layer couplers employ an intermediate waveguide layer between the upper and lower waveguides, simultaneously achieving both large layer spacing and high coupling efficiency[21]. However, this configuration requires complex fabrication processes that increase manufacturing costs.

Lithium niobate (LN) has emerged as one of the most promising platforms for PICs due to its unique combination of low absorption loss, broad transparency window, strong electro-optic (EO) effect, and high second-order nonlinearity[22-24]. Recent advances in commercially available high-quality thin-film lithium niobate (TFLN) wafers have enabled the rapid development of various TFLN-based PICs, including modulators, lasers, amplifiers, and periodically poled lithium niobate (PPLN) devices[25-30]. On the other hand, TFLN offers particular advantages for very-large-scale photonic integration (VLSPI), as its intrinsic electro-optic properties enable high-speed phase modulation without additional insertion loss while minimizing thermal dissipation during operation, significantly reducing power consumption and avoiding thermal crosstalk, both are crucial for VLSPI[31-35]. The maturation of fabrication techniques has further accelerated the development of TFLN-based PICs. Deep ultraviolet (DUV) lithography combined with optimized dry etching technology[36] has demonstrated wafer-scale fabrication of TFLN waveguides with propagation losses as low as 0.3 dB/cm. Meanwhile, the recently developed photolithography-assisted chemomechanical etching (PLACE) technique[37] enables seamless wafer-scale patterning with record-low optical losses down to 0.03 dB/cm. These technological breakthroughs have enabled the implementation of large-scale TFLN-based photonic integrated devices, with demonstrated applications spanning photonic deep learning, convolutional processors, and quantum key distribution systems[38-40]. However, the development of high-performance waveguide crossings on TFLN platforms remains scarce, with current research predominantly limited to conventional in-plane MMI-based designs. The ongoing push for VLSPI in TFLN PICs creates an urgent need for waveguide crossings with ultra-low loss and near-zero crosstalk.

In this work, we present a low-cost fabrication method for 3D interlayer couplers by exploiting the edge rounding effect inherent to chemical mechanical polishing (CMP). Combining this method with the PLACE technology, we demonstrate high-performance waveguide crossings with average loss below 0.002 dB and crosstalk below -62 dB on $Ta_2O_5$-on-LNOI integrated photonic platform. Our approach enables efficient optical coupling between vertically stacked waveguides with large interlayer spacing (1.5 μm), successfully resolves the compromise between interlayer crossing performance and coupling efficiency while eliminating the need for intermediate waveguide layer. Since the CMP technique employed in our approach is one of the fundamental semiconductor processes, the proposed method is not limited to the PLACE technology but also fully compatible with conventional semiconductor manufacturing technology.

## 2. Fabrication Methods and Design Principle

Whereas electronic integrated circuits employ vertical vias and horizontal interconnects for compact and low-loss multilayer routing[41-43], PICs face challenges in emulating this approach owing to high losses from sharp refractive index change at vertical bends. Alternatively, photonic interlayer couplers necessitate an adiabatic refractive index change to facilitate efficient optical signal propagation[44,45]. To address this challenge, we propose a method for fabricating 3D photonic interlayer couplers by leveraging the edge rounding effect inherent to CMP, as schematically presented in Figure 1. The entire fabrication process is summarized as follows. First, bottom-layer waveguides are patterned on a TFLN wafer, comprising both lateral waveguides

(to be routed upward, indicated by arrows in Figure 1) and longitudinal waveguides (oriented perpendicularly, denoted by ⊗ in Figure 1), using the PLACE technique. Detailed processes can be found in our earlier work[46]. Next, the TFLN is selectively etched at the terminal ends of the lateral waveguides via femtosecond laser ablation. Subsequently, following an RCA cleaning, a 3 μm-thick $SiO_2$ layer was deposited on the sample surface via plasma-enhanced chemical vapor deposition (PECVD). This $SiO_2$ layer is then selectively etched with a defined offset of 50 μm toward the input side of the lateral waveguides relative to the underlying TFLN etch region. Subsequently, a Ti thin film was deposited on the sample via magnetron sputtering, patterned by femtosecond laser direct writing, and utilized as a hard mask for dry etching of $SiO_2$, with an etch depth set to 1.5 μm. The sample subsequently undergoes a CMP process. During this step, protruding features (highlighted with red ellipses in Figure 1) experience higher mechanical pressure and increased abrasive interaction, leading to accelerated removal. In contrast, recessed regions (marked with purple ellipses) are removed more slowly. This differential removal behavior effectively smoothens the initially sharp topography into continuous, tapered profiles, which are essential for low-loss interlayer coupling. The CMP process continues until the TFLN within the $SiO_2$ window is completely removed, forming a gradual thickness taper. Owing to the intentional offset between the $SiO_2$ and TFLN etch steps, only the TFLN adjacent to the lateral waveguide is exposed and tapered, while the opposite side remains embedded within the $SiO_2$, thus maintaining isolation from the top-layer waveguide. Finally, a 300 nm-thick $Ta_2O_5$ layer is deposited as the top waveguide material via electron-beam evaporation, and patterned into waveguides using the PLACE technique. The 3D schematic diagram of the overall device is shown in Figure 2.

In addition to three-dimensional adiabatic routing, the optical coupling process from the bottom to the top waveguide also involves refractive index variations induced by material and structural transitions[47-49]. To maintain adiabaticity throughout the coupling region, a tapered $Ta_2O_5$ structure is fabricated in the overlap zone between the LN waveguide and the $Ta_2O_5$ top waveguide, as depicted in Figure 3(a). The figure also illustrates key cross-sectional waveguide profiles (labeled 1–4) along with their corresponding numerically simulated optical field distributions. From profiles 1 to 4, an adiabatic transition is achieved, from a pure LN ridge waveguide, through a hybrid $Ta_2O_5$–$LiNbO_3$ ridge waveguide, to a final $Ta_2O_5$ ridge waveguide. Figure 3(b) shows the simulation results of loss and crosstalk versus interlayer gap for the fabricated interlayer coupler at a wavelength of 1550 nm. The loss and crosstalk of the cross-coupled device can be observed to decrease significantly as the interlayer spacing increases. Figure 3(e) further reveals the variation curve of interlayer loss with interlayer spacing at a fixed wavelength of 1550 nm for the interlayer cross-structure. Notably, this coupler achieves low loss and low crosstalk without compromising coupling efficiency, thereby overcoming the classic performance trade-off that has long plagued traditional cross-coupler designs. For a 1.5 μm gap, Figure 3(c) simulates the wavelength dependence of interlayer coupler loss and crosstalk. Additionally, Figure 3(d) simulates the wavelength-dependent curve of interlayer cross-structure loss under a fixed gap. This demonstrates the device's ability to maintain low loss and low crosstalk across a broad wavelength range, further confirming its wide operating bandwidth.

## 3. Results

To validate the proposed fabrication approach, a test sample was fabricated comprising 20 top-layer $Ta_2O_5$ waveguides arranged with a pitch of 127 μm. Each waveguide integrates two interlayer couplers for vertical optical routing and intersects 300 underlying LN waveguides spaced at 10 μm intervals. Figure 4(a) shows a photograph of the overall sample structure. Figure 4(b) shows the transition region from the LN layer to the upper $Ta_2O_5$ waveguide. Figure 4(c) presents a magnified view of a representative interlayer coupler, while Figure 4(d) displays a detailed micrograph of the interlayer crossing region. Subsequently, the sample was sectioned using FIB. The SEM image of the cross-section of the $Ta_2O_5$-on-LNOI waveguide is shown in Figure 4(e), where the thickness of the $SiO_2$ transition layer is 1.5 μm. Figure 4(f) shows the SEM image of the pure LN waveguide cross-section in the interlayer coupler.

The performance of the fabricated device was characterized using the experimental setup illustrated in Figure 5(a). A narrow-linewidth tunable laser served as the optical source, followed by amplification via an EDFA and polarization adjustment through an in-line polarization controller. The light was then coupled into and out of the device under test (DUT) using a 3 μm ultra-high numerical aperture single-mode fiber (UHNA-7), with the input polarization set to TE mode. During measurement, the optical signals from both the through port and the cross-talk port were collected by optical fibers, converted into photocurrent using photodiodes (PDs), and recorded by an oscilloscope. Notably, a -30 dB attenuator was inserted before the PD at the through port to extend the measurable crosstalk range under the limited responsivity of the detector. By calculating the optical signal data collected from both ports, the insertion loss at each individual waveguide crossing within this layer coupler was obtained, as shown in Figure 5(b). Figure 5(c) and 5(d) show the measured through-port crosstalk as a function of wavelength when the laser propagated through the bottom LN waveguide and the top $Ta_2O_5$ waveguide, respectively. The results demonstrate high-performance waveguide crossings in both cases, with an average loss below 0.002 dB and crosstalk lower than -62 dB across the measured band.

Table 1 lists the measurement results of waveguide crossings fabricated on different platforms in recent years, comparing three key metrics: wavelength range, insertion loss, and crosstalk. The wavelength range directly determines the operating bandwidth of waveguide crossings devices in broadband applications, making it a key metric for assessing their spectral flexibility. In Table 1, LNOI devices exhibit a more moderate wavelength coverage range (typically 40–180 nm), with specific values varying depending on their design parameters. Although the LNOI device[6] exhibits the widest wavelength span (1500 nm to 1680 nm, 180 nm), this advantage comes with relatively high insertion loss (0.12 dB) and crosstalk (-40 dB); With an extremely low insertion loss of 0.043 dB (in the 1550-1560 nm wavelength range), silicon emerges[12] as the material with the most advantageous overall performance in Table 1. Silicon's high refractive index (n=3.48) is key to achieving strong mode confinement, which effectively suppresses light field leakage and reduces waveguide loss to an extremely low level. This is precisely why it delivers outstanding performance. However, silicon's narrow spectral range limits its practicality in applications requiring broadband operation, as it achieves excellent low loss but has limited wavelength coverage capability. The refractive index of SiN (~1.99) is comparable to that of lithium niobate thin films, enabling device designs with similar dimensions and performance characteristics[8,9]. The insertion loss values for LNOI and SiN devices span a wider range, from 0.07 dB to 0.6 dB, depending on the specific design and wavelength span under consideration; Crosstalk is a key parameter for evaluating waveguide cross-performance, as it fundamentally reflects the level of undesirable coupling between adjacent waveguides. Devices based on Si, SiN, and LNOI all achieve exceptional crosstalk suppression (-31 dB to -50 dB), with this metric varying depending on device design and operating conditions. Based on the performance comparison of the aforementioned devices, the 3D waveguide crossing we developed demonstrates outstanding performance: it achieves a wavelength coverage of 120 nm (from 1510 nm to 1630 nm) while simultaneously maintaining extremely low insertion loss (0.02 dB) and crosstalk (-62 dB). These outstanding performance metrics position the device as one of the most advanced designs reported for lithium niobate waveguide crossings, effectively meeting the diverse demands of broadband high-performance photonic systems and demonstrating broad application prospects. In the future, we will develop various high-performance photonic integrated chips based on the $Ta_2O_5$-on-LNOI platform, such as optical switches and photonic neural network chips.

## 4. Conclusion

In conclusion, we have developed and demonstrated a novel, cost-effective fabrication method for high-performance 3D waveguide crossings on the $Ta_2O_5$-on-LNOI platform by leveraging the edge rounding effect inherent in the CMP process. Our approach successfully overcomes the long-standing trade-off between interlayer crossing loss and coupling efficiency, achieving record-low insertion loss (<0.002 dB) and crosstalk (<–62 dB) without requiring intermediate waveguide layers or complex multi-step coupling structures. The proposed technique is fully compatible with conventional semiconductor manufacturing processes. Our results open a practical route toward VLSPI of TFLN devices for applications in high-speed communications, optical computing, and quantum information processing.

# Tables

Table 1. Performance Comparison of Waveguide Crossing

| Platfrom | Wavelength Range (nm) | Wavelength Span (nm) | Insertion Loss (dB) | Crosstalk (dB) | References |
|---|---|---|---|---|---|
| Si | 1540-1560 | 20 | 0.2 | -40 | [10] |
| Si | 1500-1600 | 100 | 0.072 | -33 | [7] |
| Si | 1550-1560 | 10 | 0.043 | -50 | [12] |
| SiN | 1520-1640 | 120 | 0.6 | -42 | [9] |
| SiN | 1540-1560 | 20 | 0.08 | -44 | [8] |
| LNOI | 1500-1600 | 100 | 0.48 | -36 | [22] |
| LNOI | 1500-1680 | 180 | 0.12 | -40 | [6] |
| LNOI | 1520-1600 | 80 | 0.07 | -50 | [21] |
| LNOI | 1540-1580 | 40 | 0.09 | -31 | [11] |
| LNOI | 1510-1630 | 120 | 0.002 | -62 | This work |

# Figures

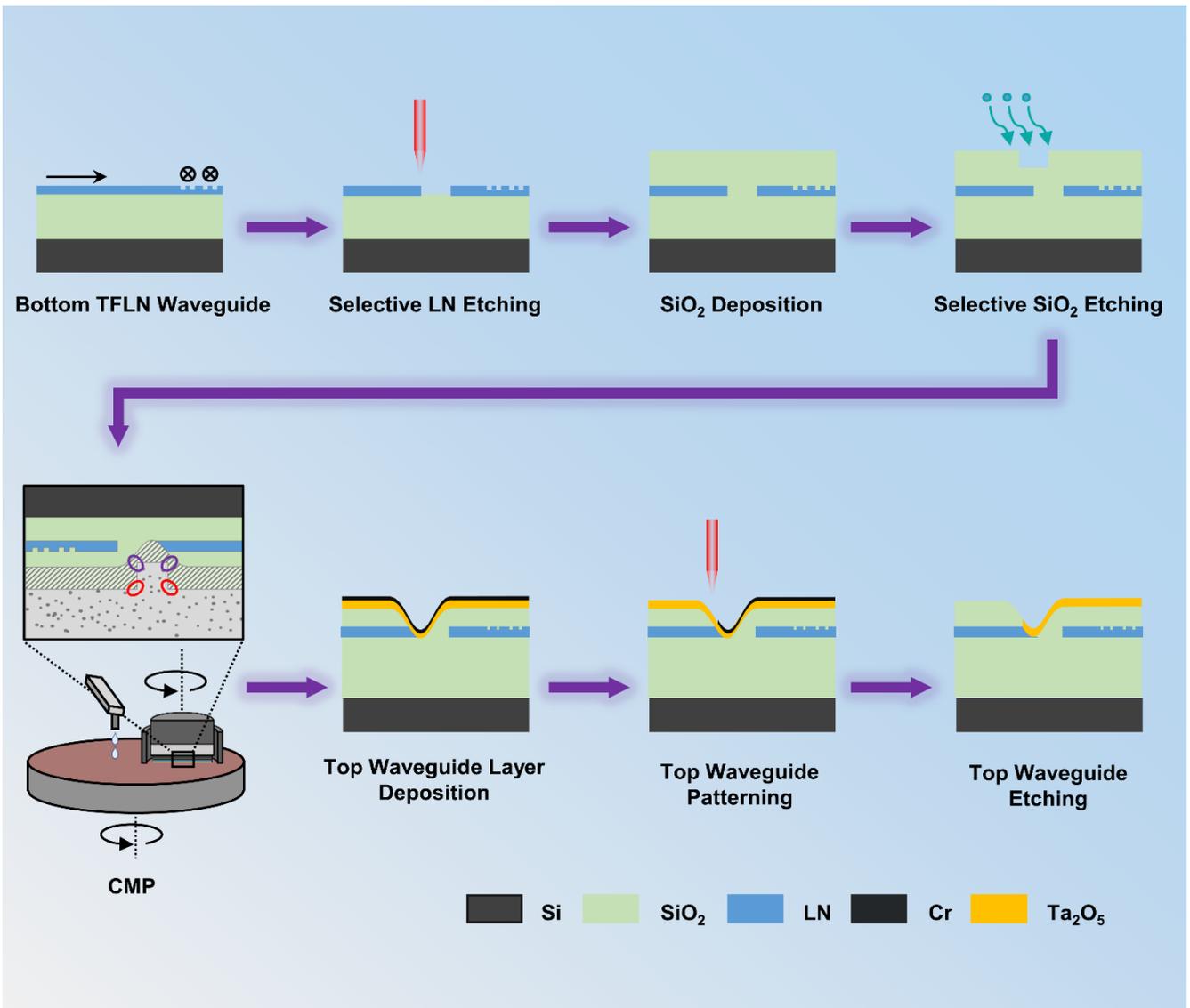

**Figure 1.** Schematic fabrication flow of the proposed 3D interlayer coupler enabling adiabatic light transition between vertically stacked waveguides.

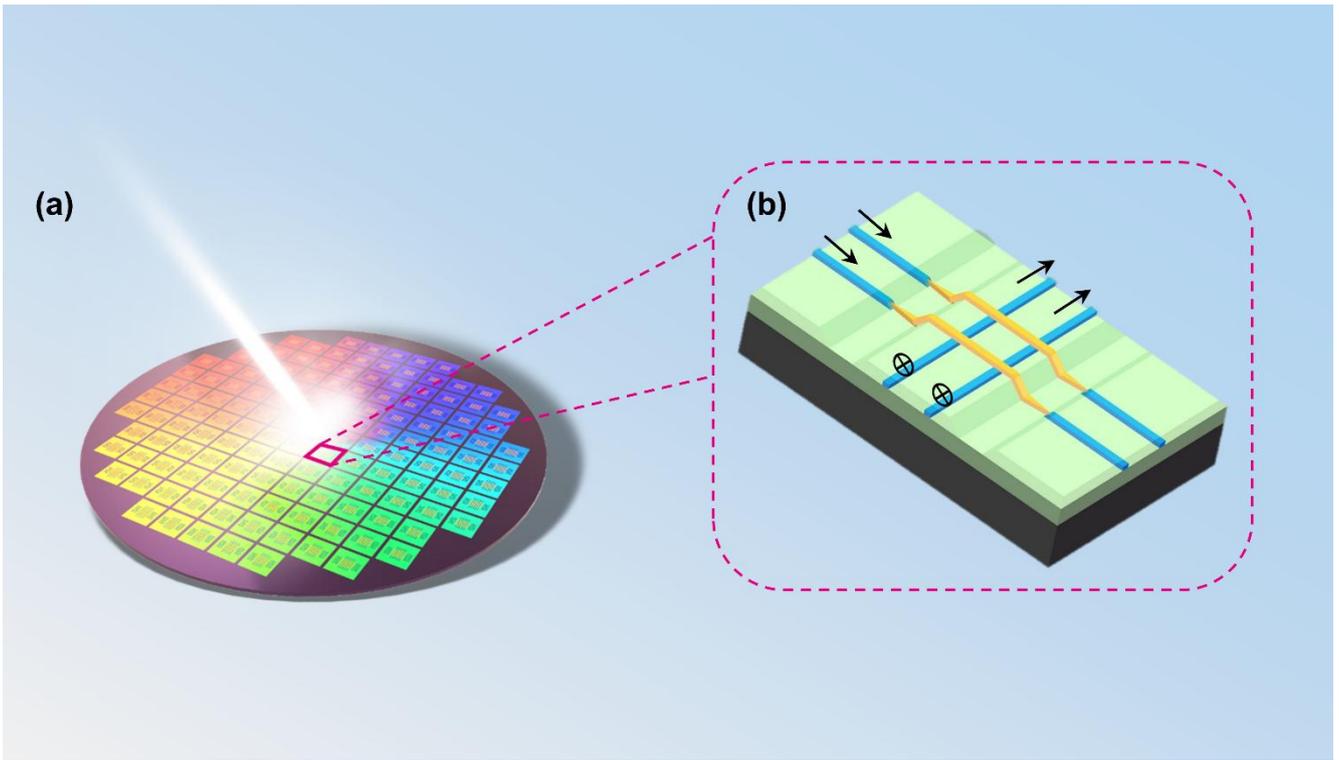

**Figure 2.** 3D device schematic for Ta$_2$O$_5$-on-LNOI interlayer coupler. (a) 3D assembly diagram of the fabricated device. (b) 3D close-up view of a single interlayer coupler.

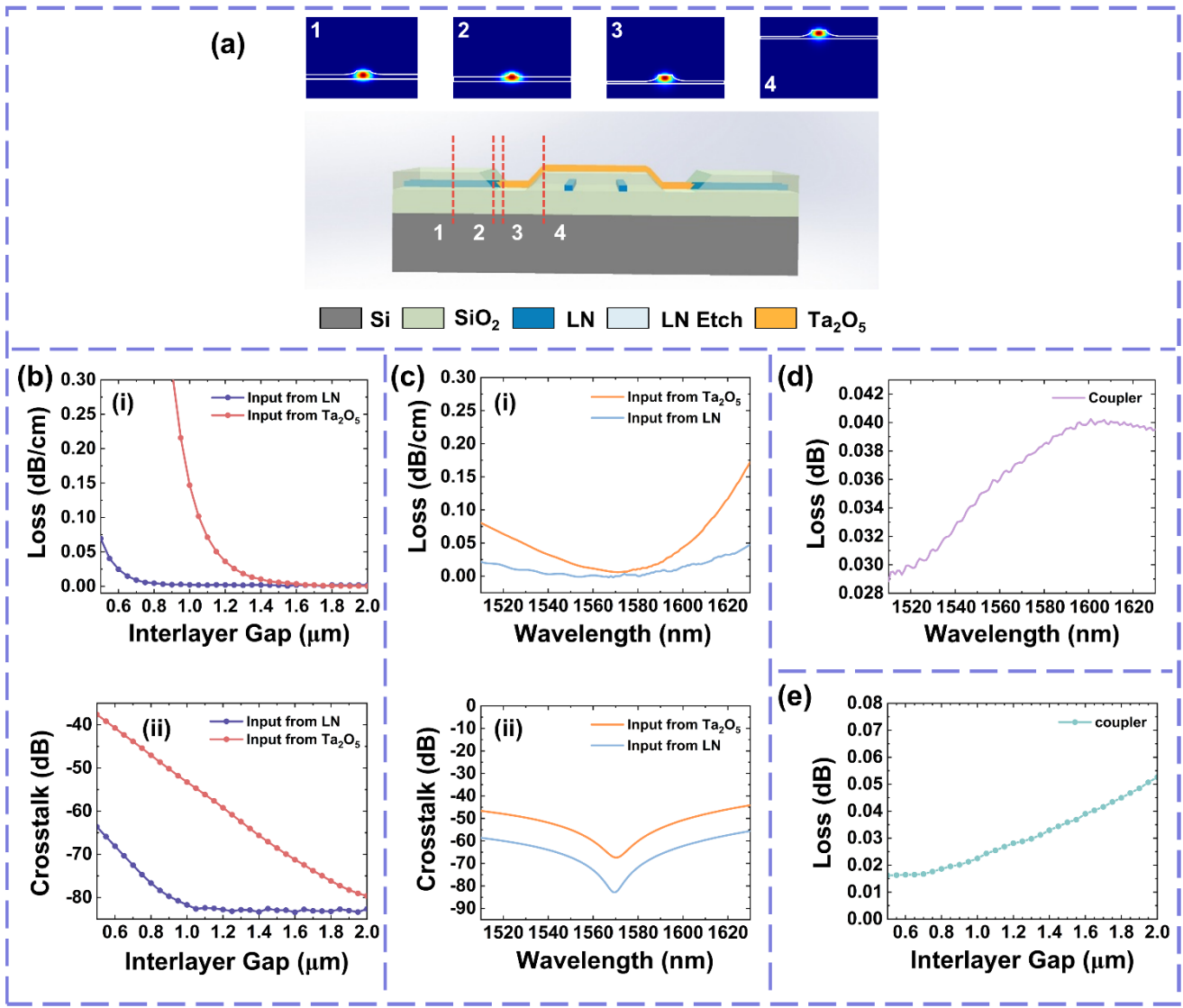

**Figure 3.** Optical simulation results for Ta$_2$O$_5$-on-LNOI interlayer coupler. (a) 3D schematic of the interlayer coupler and cross-sectional optical field evolution (profiles 1–4). (b) Simulated loss (i) and crosstalk (ii) of the interlayer coupler versus interlayer gap at 1550 nm wavelength. (c) Simulated wavelength-dependent loss (i) and crosstalk (ii) at 1.5 μm interlayer gap for interlayer coupler. (d) Simulated loss of the interlayer crossing versus interlayer gap at 1550 nm wavelength. (e) Simulated wavelength-dependent loss at 1.5 μm interlayer gap for the interlayer crossing.

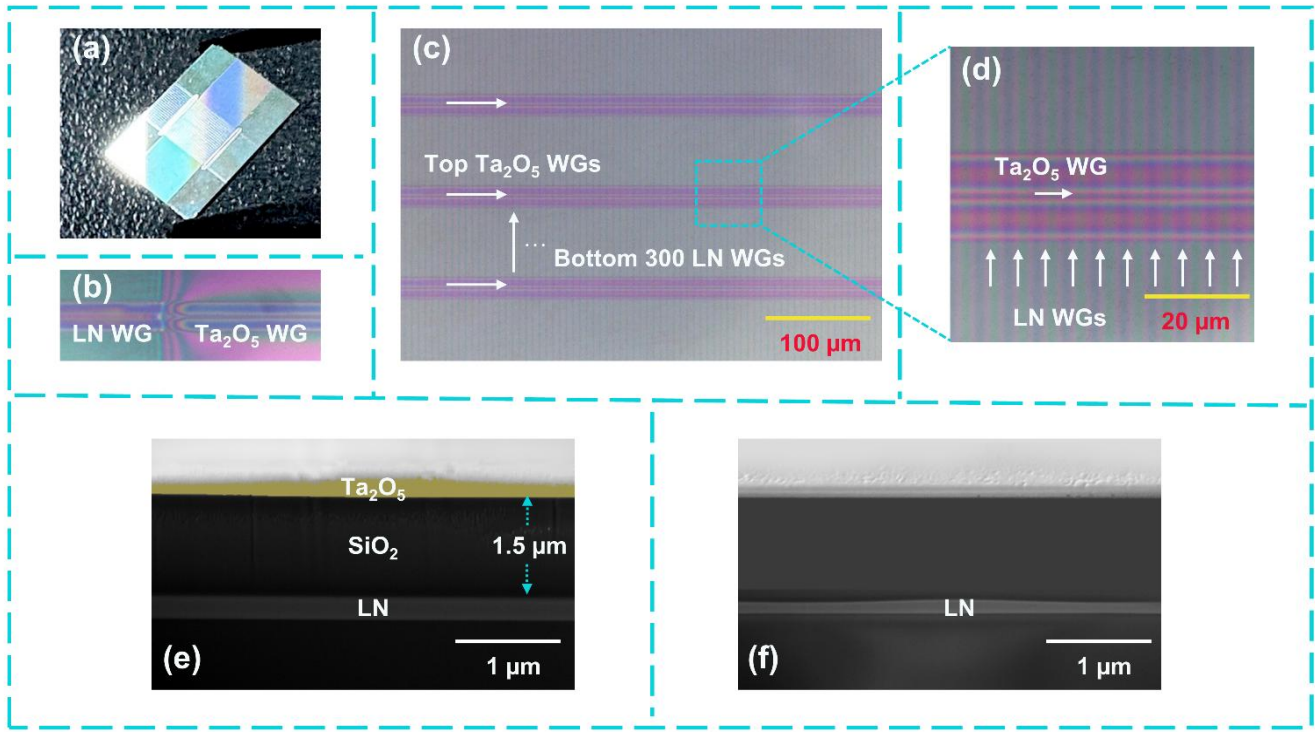

**Figure 4.** Optical microscope image of Ta₂O₅-on-LNOI integrated photonic device. (a) Photograph of the overall device with 20 top Ta₂O₅ waveguides (pitch: 127 μm). (b) Taper transition region connecting the bottom LN waveguide layer to the top Ta₂O₅ waveguide layer. (c) Magnified view of a single interlayer coupler. (d) Close-up of the interlayer crossing region, showing intersection with underlying LN waveguides. (e) SEM image of the Ta₂O₅-on-LNOI waveguide cross-section, where the SiO₂ transition layer has a thickness of 1.5 μm. (f) SEM image of the bottom LN layer waveguide cross-section.

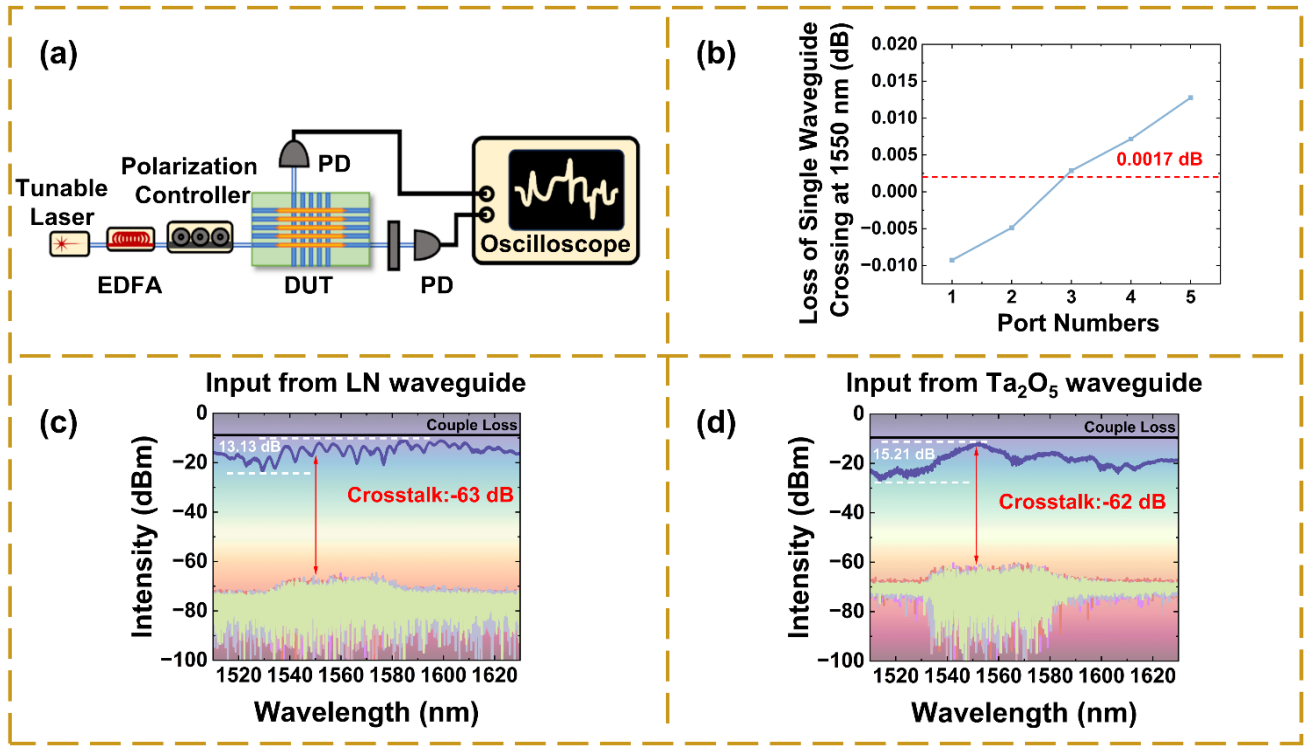

**Figure 5.** Experimental setup and performance characterization of waveguide crossings. (a) Schematic of the optical characterization system using a tunable laser, EDFA, polarization controller, UHNA-7 fiber, and attenuated detection. (b, c, d) Wavelength-dependent through loss and crosstalk for light propagating through the bottom LN and top $Ta_2O_5$ waveguides, respectively, showing <0.002 dB loss and <–62 dB crosstalk.